\documentclass{webofc}
\usepackage[varg]{txfonts}   % Web of Conferences font
\usepackage{bm}
\usepackage{braket}
\begin{document}
\title{Analysis of coupled-channel potentials with quark and hadron degrees of freedom}

\author{
    \firstname{Ibuki} \lastname{Terashima}\inst{1}\fnsep\thanks{\email{terashima-ibuki@ed.tmu.ac.jp}} \and
    \firstname{Tetsuo} \lastname{Hyodo}\inst{1}\fnsep\thanks{\email{hyodo@tmu.ac.jp}}
    }

\institute{Department of Physics, Tokyo Metropolitan University, Hachioji 192-0397, Japan}

\abstract{%
The quark-antiquark potentials are known to be confining in the absence of the $\bar{q}q$ pair creation. On the other hand, the inter-hadron potentials vanish at large distance, because the interaction range is limited by the inverse pion mass. When the $\bar{q}q$ pair creation and annihilation are switched on, the channel coupling occurs between the quark and hadron degrees of freedom, modifying the behavior of potentials. In this work, we investigate the properties of the effective potentials obtained by eliminating one of the channels. We show that the coupling to the eliminated channel induces a non-local and energy dependent effective potential, irrespective of the properties of the transition potential. In addition, when the hadron channel having continuous scattering eigenstates is eliminated, the resulting inter-quark potential contains an imaginary part above the threshold to describe the decay effect into the meson pair.
}
\maketitle
\section{Introduction}
\label{intro}

% Potentials in hadron physics
Potential descriptions in the strong interaction phenomena are sometimes useful, even though the fundamental theory of QCD is formulated as a quantum field theory. For instance, quark-antiquark $(\bar{Q}Q)$ potentials in the static limit can be used to study the color confinement~\cite{Wilson:1974sk}, and the nuclear force potentials are the basis to study various few-body and many-body physics of the atomic nuclei~\cite{Epelbaum:2008ga}. Recent developments of lattice QCD calculations~\cite{Bali:2000gf} enable us to study these potentials from the first principles.

% quark-hadron coupling, finite quark mass
The potentials with the quark degrees of freedom and those with hadron degrees of freedom are related with each other, through the creation/annihilation of a $\bar{q}q$ pair. In fact, the static $\bar{Q}Q$ potential shows the ``string breaking'' indicated by the flattening of the potential at large distance~\cite{Bali:2005fu}. This behavior should be qualitatively modified for the case with finite quark masses, because of the existence of the continuum of the meson-meson states. The lattice QCD studies of the $\bar{Q}Q$ potential with finite quark masses have been performed in Refs.~\cite{Ikeda:2011bs}, but the effect of the meson-meson thresholds is not well understood. 

% This work
In this work, we consider the channel coupling between the quark and hadron degrees of freedom. By constructing a coupled-channel potential model, we discuss the effect of the meson-meson states for the $\bar{Q}Q$ potential with finite quark masses. At the same time, our formulation clarifies the effect of the quark dynamics in the hadron potentials. Such analysis will be important to understand the nature of the $H$ dibaryon in the strangeness $S=-2$ baryon-baryon scattering, where the baryon-baryon potential may contain the effect of the six-quark state generated by the quark potential.

\section{Formulation}

% Hamiltonian
For concreteness, we consider the coupled system of the quark-antiquark channel (denoted as $\bar{c}c$) and hadron-hadron channel ($\bar{D}D$) described by the Schr\"odinger equation
\begin{equation}\label{hamiltonian}
    H\ket{\Psi}
    =E\ket{\Psi} , \quad
    H=
    \begin{pmatrix}
    T^{\bar{c}c} & 0 \\
    0 &  T^{\bar{D}D}+ \Delta\\
    \end{pmatrix}
    +
    \begin{pmatrix}
    V^{\rm conf} & V^t \\
    V^t  & V^{\bar{D}D} \\
    \end{pmatrix},
    \quad
    \ket{\Psi}
    =
    \begin{pmatrix}
    \ket{ \psi^{\bar{c}c} }\\
	\ket{    \psi^{\bar{D}D}} 
    \\
    \end{pmatrix},
\end{equation}
where $T^{\bar{c}c}$ ($T^{\bar{D}D}$) is the kinetic energy of the $\bar{c}c$ ($\bar{D}D$) channel and $\Delta$ is the threshold energy of $\bar{D}D$. Here all the potentials are assumed to be local ones, i.e., $\bra{\bm{r}^{\prime}}V\ket{\bm{r}}=V(\bm{r})\delta(\bm{r}^{\prime}-\bm{r})$. The diagonal potential in the $\bar{c}c$ channel $V^{\rm conf}$ is confining, and the $\bar{D}D$ potential $V^{\bar{D}D}$ vanishes at large distance. The channel coupling is induced by the transition potential $V^t$.

% effective potentials
Following the Feshbach method~\cite{Feshbach:1958nx}, we can construct the single-channel effective Hamiltonian for the $\bar{c}c$ channel $H^{\bar{c}c}_{\rm eff}(E)$ as
\begin{equation}
    H^{\bar{c}c}_{\rm eff}(E) = T^{\bar{c}c}+V^{\rm conf}+V^t
    (E-T^{\bar{D}D}- \Delta-V^{\bar{D}D}+i0^+)^{-1}
    V^t \equiv T^{\bar{c}c}+ V^{\bar{c}c}_{\rm eff}(E),
\end{equation}
with $H^{\bar{c}c}_{\rm eff}(E)\ket{\psi^{\bar{c}c}}=E\ket{\psi^{\bar{c}c}}$. The coupling with the $\bar{D}D$ channel is renormalized in the effective potential $V^{\bar{c}c}_{\rm eff}(E)$. We derive, in the same way, the effective Hamiltonian for $\bar{D}D$ as
\begin{equation}
    H^{\bar{D}D}_{\rm eff}(E) = T^{\bar{D}D}+\Delta+V^{\bar{D}D}+V^t(E-T^{\bar{c}c}-V^{\rm conf}+i0^+)^{-1}V^t \equiv T^{\bar{D}D}+\Delta+ V^{\bar{D}D}_{\rm eff}(E),
\end{equation}
with $H^{\bar{D}D}_{\rm eff}(E)\ket{\psi^{\bar{D}D}}=E\ket{\psi^{\bar{D}D}}$. By solving the single-channel effective Schr\"odinger equations in a self-consistent manner for the energy $E$, we obtain the results equivalent to the original coupled-channel equation~\eqref{hamiltonian}.

\section{Results}

% DDbar potential
We first investigate the coordinate representation of the effective potential of the $\bar{D}D$ channel,
\begin{equation}\label{VDD}
    \bra{ \bm{r}'_{\bar{D}D}}
    V^{\bar{D}D}_{\rm eff}(E)
    \ket{\bm{r}_{\bar{D}D}}
    =V^{\bar{D}D}(\bm{r})\delta( \bm{r}'- \bm{r}) 
    +\sum_n \frac{\bra{\bm{r}'_{\bar{D}D} } V^{t}\ket{\phi_n }
    \bra{\phi_n} V^{t} \ket{\bm{r}_{\bar{D}D} }}{E-E_n} ,
\end{equation}
where $\ket{\phi_n}$ and $E_{n}$ respectively represent the eigenstate and eigenenergy of the $\bar{c}c$ channel without the channel coupling, $(T^{\bar{c}c}+V^{\rm conf})\ket{\phi_n}=E_{n}\ket{\phi_n}$. Because of the confinement, $\ket{\phi_n}$ consists only of the discrete eigenstates. The  second term shows that the effective potential is a non-local one, because the $\bm{r}$ dependence is factorized from that of $\bm{r}^{\prime}$. This means that a non-local potential is induced by the channel coupling regardless of the property of the transition potential $V^t$. In addition, the denominator of the second term shows that the effective potential depends on the energy $E$. It can also be seen that the potential strength diverges at $E=E_{n}$ when the energy coincides with an eigenenergy of $\ket{\phi_n}$.

% ccbar potential
Next, we consider the effective $\bar{c}c$ potential. Here we assume that the $\bar{D}D$ potential $V^{\bar{D}D}$ does not support any bound states, so that the eigenstates without the channel coupling are given only by the continuum of the scattering states labeled by $\bm{p}$: $(T^{\bar{D}D}- \Delta-V^{\bar{D}D})\ket{\bm{p}}=E_{\bm{p}}\ket{\bm{p}}$. The lowest energy state with $\bm{p}=\bm{0}$ appears at the threshed, $E_{\bm{0}}=\Delta$. In this case, the coordinate representation of the effective $\bar{c}c$ potential is 
\begin{equation}\label{Vccexp}
    \bra{ \bm{r}'_{\bar{c}c}}
    V^{\bar{c}c}_{\rm eff}(E)
    \ket{\bm{r}_{\bar{c}c}}
    =V^{\rm conf}(\bm{r})\delta( \bm{r}'- \bm{r}) 
    +\int d\bm{p} \frac{\bra{\bm{r}'_{\bar{c}c} } V^{t}\ket{\bm{p}}
    \bra{\bm{p}} V^{t} \ket{\bm{r}_{\bar{c}c} }}{E-E_{\bm{p}}+i0^{+}} .
\end{equation}
As in Eq.~\eqref{VDD}, the effective potential is non-local and energy-dependent. For the energy $E\geq \Delta$, the potential has an imaginary part due to the pole at $E = E_{\boldsymbol{p}}$, which is responsible for the opening of the $\bar{D}D$ threshold, above which the decay process $\bar{c}c\to \bar{D}D$ occurs.

% analytic form
In the absence of the diagonal $\bar{D}D$ potential ($V^{\bar{D}D}=0$), we can perform the integration in Eq.~\eqref{Vccexp} analytically to obtain 
\begin{equation}\label{Vcccond}
    \bra{ \bm{r}'_{\bar{c}c}}
    V^{\bar{c}c}_{\rm eff}(E)
    \ket{\bm{r}_{\bar{c}c}}
    = V^{\rm conf}(\bm{r})\delta(\bm{r}'-\bm{r}) 
    -V^t(\bm{r}')V^t(\bm{r})
    \frac{m_D\exp[-\sqrt{m_D(\Delta - E)-i0^+}
    |\bm{r}'-\bm{r}|]}{4\pi|\bm{r}'-\bm{r}|} ,
\end{equation}
with $m_{D}$ being the mass of the $D$ meson. In this expression, the appearance of the imaginary part for $E \geq \Delta$ is represented as $\sqrt{m_D(\Delta - E)-i0^+}=-i\sqrt{m_D(E-\Delta)}$. Because of the $V^t(\bm{r}')V^t(\bm{r})$ factor, this is still a non-local potential. By performing the derivative expansion, we obtain the local approximation
\begin{equation}
    \bra{ \bm{r}'_{\bar{c}c}}
    V^{\bar{c}c}_{\rm eff}(E)
    \ket{\bm{r}_{\bar{c}c}}
    =
    V^{\rm conf}(\bm{r})\delta(\bm{r}'-\bm{r}) 
    +\frac{[V^t(\bm{r})]^2}{E-\Delta }\delta(\boldsymbol{r'}-\boldsymbol{r})+O(\nabla^2).
\end{equation}
This expression, however, has no imaginary part for $E\geq \Delta$, in  contrast to Eq.~\eqref{Vcccond}. We thus find that a naive derivative expansion of the non-local potential may break the physical property of the potential.

\section{Conclusion and future outlook}

% summary
We have discussed the effect of the transition between the quark channel with a confining potential and the hadron channel with a scattering potential. It is shown that the channel coupling introduces the non-locality and energy-dependence in the effective single-channel potentials. We find that the coupling to the hadron channel with continuous eigenstates generates the imaginary part of the effective potential above the threshold, which however, disappears when the formal derivative expansion is applied to obtain a local potential. This indicates that the construction of a local effective potential should be performed with care.

% future outlook
As a future outlook, it will be interesting to examine various local approximations other than the formal derivative expansion. For instance, an alternative method to obtain a local potential is proposed in Ref.~\cite{Aoki:2021ahj}. Another direction is to apply the present formulation to the actual hadron systems. The charmonium like $X(3872)$ can be described as the $D\bar{D}^{*}$ scattering coupled with the $\bar{c}c$ core, which is a suitable system for an application.

%\bibliography{refs.bib}

%\begin{thebibliography}{}
%%
%% and use \bibitem to create references.
%%
%
%\bibitem{s-b} G.S. Bali et al., Phys. Rev. D \textbf{71}, 114513 (2005).
%\bibitem{qq} Y. Ikeda, H. Iida, Prog. Theor. Phys. \textbf{128}, 941 (2012);T. Kawanai, S. Sasaki, Phys. Rev. Lett. 107, 091601 (2011).
%\bibitem{had} N.~Ishii, S.~Aoki and T.~Hatsuda,
%%``The Nuclear Force from Lattice QCD,''
%Phys. Rev. Lett. \textbf{99}, 022001 (2007);Y. Akahoshi, et al., PTEP \textbf{2020}, 073B07 (2020).
%\bibitem{fesh}
%H. Feshbach, Ann. Phys. \textbf{5}, 357 (1958); ibid., \textbf{19}, 287 (1962)
%\end{thebibliography}

\end{document}